\documentstyle[prl,aps,multicol,epsfig]{revtex}

\begin{document}

\draft

\title{\quad \quad \quad { comment on "Two Fermi points in the exact solution of the
Kondo problem"  }}

\author{ N. Andrei}

\address{Center for Materials Theory, Serin Physics Laboratory, Rutgers University\\
                        Piscataway, New Jersey 08854-8019, USA}
\date{December 8, 2001}
\maketitle

\begin{multicols}{2}

\narrowtext

 In a recent paper \cite{zvyagin} A. Zvyagin reexamined the solution 
of the Kondo model, and claimed that the model should be considered 
with two "Fermi points"
 in the presence of a magnetic field, and that a new energy 
scale arises as a result. I show below that these claims are in error. 

\bigskip

The Kondo hamiltonian,
$$
H = -i \int dx ~\psi^{\dagger}_a(x) \partial  \psi_a(x) + 
J \psi^{\dagger}_a(0) \sigma^i_{ab}  \psi_b(0)
$$
describes a magnetic impurity in a metal, capturing
 the radial dynamics of a 3d model.  The spectrum is linearized
 around the Fermi point (there is only 1 point since the Fermi surface 
in 3d is simply connected) and as a result the electrons are chiral - right
movers by our convention - with a spectrum $E= p$.

 The linearization is valid for $J$ small and for momenta small 
compared to a cutoff $D$, of the order of the bandwidth. 
The cutoff is considered large compared to any physical quantity
and the only scale remaining in the problem 
is the Kondo temperature $T_K = D \exp^{- \frac{\pi}{2J}}$.
 The results are universal and parameterized by $T_K \ll D$ . If the
 coupling is not small the spectrum cannot be 
linearized and the universality is lost.

The model is soluble
and its solution is  given by the set of equations:
\begin{equation} \label{bae}
N^e \theta_1(\lambda_{\alpha}-1) + \theta_{1}(\lambda_{\alpha} ) =
2\pi J_{\alpha} +\sum_{\beta =1}^M \theta_2(\lambda_{\alpha}- \lambda_{\beta})
\end{equation}
where $\theta_{n}(\lambda) = 
- 2 \tan^{-1}(2\lambda /cn)$, $c= 2J/(1-3/4 J^2)$, $N^e$ is the number of electrons interacting with a spin 1/2 impurity.
 There is a total of $N =N^e+1$ spins 
(including the impurity), $M$  of which are "down" and $N-M$ are "up".
The spin of the state is $S= \frac 1 2 (N-2M)$.
 The $M$  integers (or half integers)  $J_{\alpha}$ 
are the quantum numbers of the states and each allowed configuration
of them determines the state uniquely. From equation (\ref{bae}) we find
that the  allowed values of $M$  quantum numbers $J_{\alpha}$ are between 
 $J_{max} =(N-M-1)/2$ and $J_{min}= -J_{max}$. Having solved the equations
for a {\em distinct} set of $\lambda$'s the energy of the state is given by
$$
E= D \sum_{\alpha=1}^M [\theta_1(\lambda_{\alpha}-1) - \pi ]  + \sum_{j=1}^{N^e} \frac{2 \pi}{L}  n_j
$$
where $L$ is the size of the system, $D=N^e/L$ is the electron density (also serving as a cut off) and $n_j$ are integers, the charge quantum numbers of the system.

The ground state of the system is a singlet $S=0$ with the $M=N/2$ 
 quantum numbers  $J_{\alpha}$ occupying consecutively all  
$N/2$ slots between 
$J_{max}=(N/2-1)/2$ and $J_{min}=-(N/2-1)/2$. In the thermodynamic limit
$N,M \to \infty$ and also $\lambda_{\alpha}$ takes values from $-\infty$ to 
$\infty$. We shall denote the density of solutions by $\sigma(\lambda)$.
Excited states 
correspond to other choices
of allowed configurations. The simplest is the triplet, with
$M= N/2 -1$, leaving two "holes" in the sequence of quantum numbers between
$J_{max}=(N/2)$ and $J_{min}= -J_{max}$. Each of these holes describes a spin-1/2 spinon with energy \cite{and},
\begin{equation}\label{sp}
 E^{spinon} =2D~ \tan^{-1} \left[ \exp \frac{\pi}{c}(\lambda ^{h} - 1) \right]
\end{equation}
where $\lambda ^{h}$ is the spin-rapidity of the spinon, corresponding to the
unfilled slot $J^h$. The hole can take any value in the allowed range
 and correspondingly $\lambda ^{h}$ can take any real value in the 
thermodynamic limit. Note that for $\lambda ^{h}$ large and negative 
the energy is arbitrarily small while for $\lambda ^{h}$ large and 
positive it  very large, of the order of the cutoff. This asymmetry of 
the spinon spectrum is inherited from the asymmetry inherent in
the linear spectrum of the  right moving electrons. The spectrum of the
spinons becomes linear in the momentum when the cutoff is removed \cite{nat}.
(This asymmetry is in clear 
contradiction to statements in the last paragraph of the 
first column on page 060405 of ref. \cite{zvyagin}.)

A typical excited state will have 
holes (namely spinons) and may also have "strings" (complex $\lambda$'s)
combining the spinons to lower spin states. For example, in 
addition to the  two spinon triplet, there is a two spinon 
singlet, degenerate in energy with the triplet in the 
infinite volume limit, consisting of two holes and a 2-string.

 We now turn on a magnetic field $h$ at zero temperature, adding to the hamiltonian the term
\begin{equation}
\nonumber
H_{mag} = -2S~h,
\end{equation}
where S is the total spin component of the system in the direction of $h$. Note that $H_{mag}$ commutes with the hamiltonian.

  As a result of the magnetic term, the system will gain energy by 
flipping spins to align with the magnetic field.  Each flipped spin 
corresponds to two holes in the $\lambda$ sea, two
 spinons.  The excitation energy of each spinon is given by Eq.(\ref{sp}),
and  can be made arbitrarily close to zero by choosing 
$\lambda$$^{h}$ sufficiently large and negative. Hence the magnetic field will excite spinons with $\lambda$ large and negative. As 
all $\lambda$ rapidities must be distinct, we are 
led to a depletion region where no $\lambda$ solutions exist from 
$\lambda$ =  $- \infty$ to $\lambda$ = B. 

The ``magnetic Fermi point''
$B$ 
is determined by an equilibrium argument equating the magnetic 
energy gained by flipping spins to align with the magnetic field and 
the energy cost of these holes. One finds

$$
B = B(h) = \frac{c}{\pi} \ln( \frac{h}{T_h})
$$
With $T_h$ a magnetic Kondo scale.
The main point we wish to emphasize is that  due to the form of the
spinon excitation spectrum there is only one depletion region, not two as 
claimed by Zvyagin. 

It is  obvious that no complex strings are excited, as these reduce the 
total spin of the system.  For example, consider the fundamental singlet 
and triplet excitations.  While their cost in interaction energy is 
the same, the singlet does not gain magnetic energy.
We thus have to consider the following magnetization 
equation determining the lowest energy state in the presence of a magnetic field \cite{and,wieg,rev}:
\begin{equation} \label{mag}
\sigma_{B}(\lambda)+ \int^{\infty}_{B} K(\lambda - \lambda')
\sigma_{B}(\lambda')d\lambda'=f_{kondo}(\lambda),
\end{equation}
where, 
\begin{eqnarray}
K(\lambda) &= & \frac{1}{\pi}\frac{c}{c^{2}+\lambda^{2}} = \frac{1}{\pi} 
\frac{d}{d\lambda} \theta(\lambda) \nonumber \\
f_{kondo}(\lambda)&= & \frac{N^{e}}{\pi} \frac{(c/2)}{(c/2)^{2}+(\lambda-1)^{2}}+\frac{1}{\pi}\frac{(c/2)}{(c/2)^{2}+\lambda^{2}}. \nonumber
\end{eqnarray}
The energy and total spin of the system are given by
\begin{eqnarray}
E(h)= E_{B(h)}&=&D \int^{\infty}_{B} \sigma_{B}(\lambda)[\theta_1(\lambda -1)-\pi] \nonumber \\
&+&\sum_{j} \frac{2\pi}{L} n_{j} -2h~S
\end{eqnarray}
and 
\begin{equation}
S=\frac{1}{2}N -M =\frac{1}{2}N - \int^{\infty}_{B} \sigma_{B}(\lambda)d\lambda.
\end{equation}
The quantum numbers ${n_{j}}$ have no spin content and are not excited by the magnetic field.

Equation (\ref{mag}) is a Wiener-Hopf integral equation and has been discussed using 
this technique by Yang and Yang in connection with a similar problem in the 
Heisenberg model \cite{yy}.  It is amusing to point out that while Eq. (\ref{mag}) is exact 
in the case of the Kondo model, it is only approximate in the Heisenberg case.  
The reason is simple.  The excitation spectrum for the Heisenberg model,
$$
 E^{spinon} = \frac{2J}{ \cosh(\frac{\pi}{2} \lambda^h)}
$$
so that holes will be excited in the low energy regions  
$(- \infty~, -B]$ and $[B, + \infty)$, and solutions $\lambda_{\alpha}$ will
fall into the segment $[-B, B]$. Hence  the magnetization equation 
takes the form \cite{yy},
$$
\sigma_{B}(\lambda)+ \int^{B}_{-B} K(\lambda - \lambda')
\sigma_{B}(\Lambda')d\lambda'=f_{heis}(\lambda),
$$
where
$$
f_{heis}(\lambda)= \frac{N}{\pi} \frac{(c/2)}{(c/2)^{2}+(\lambda)^{2}}
$$
As this equation is not of the Wiener Hopf form Yang and Yang
develop a method to take into account both depletion regions 
for small magnetic fields (large $B$).

  As another example consider  the closely related
  Gross-Neveu model. The spinon spectrum is,
$$ 
E^{spinon} = m \cosh (\frac{\pi}{c} 
\lambda^h)
$$
and holes will be excited in the low energy region   $[-B,~B]$. Solutions $\lambda_{\alpha}$ 
exist therefore only in the regions $(-\infty, -B]$ and $[B, \infty)$
and we are led to the magnetization equation,

$$
\sigma_{B}(\lambda)+ \left( \int_{-\infty}^{-B}+ \int_{B}^{\infty}\right) K(\lambda - \lambda')
\sigma_{B}(\Lambda')d\lambda'=f_{GN}(\lambda),
$$
with
$$
 f_{GN}(\lambda)= \frac{N^{+}}{\pi} \frac{(c/2)}{(c/2)^{2}+(\lambda-1)^{2}}+
\frac{N^-}{\pi}\frac{(c/2)}{(c/2)^{2}+(\lambda+1)^{2}}.
$$

To summarize, the form of the magnetization equation depends on the spectrum of excitations. While for the Kondo model we have only one depletion 
region, $(-\infty~,B]$
and only {\em one} magnetic Fermi point in the Kondo model, there are 
two points in the other models due to the symmetric form of their excitation spectrum.  

 In his  paper A. Zvyagin has introduced  a second depletion region 
and second Fermi point to the Kondo magnetization equation. The state thus constructed has
excitations in the range $[B, ~\infty)$ with each hole having energy 
of the order of the cutoff. The state is therefore
  infinitely excited  and has 
no meaning.

\end{multicols}
\end{document}